\documentclass[11pt,letterpaper]{article}
\usepackage[utf8]{inputenc}
\usepackage[USenglish,spanish]{babel}
\usepackage{amsmath}
\usepackage{amsfonts}
\usepackage{amssymb}
\usepackage{graphicx}
\usepackage[left=3cm,right=3cm,top=2cm,bottom=3cm]{geometry}

\usepackage[small,bf]{caption} 

\author{Jorge Pinochet}
\title{\textbf{Explorando los agujeros negros\\ 
\large Exploring black holes}}

\begin{document}

\renewcommand{\figurename}{\textbf{Figura}} 

\author{Jorge Pinochet$^{*}$\\ \\
 \small{$^{*}$\textit{Departamento de Física, Universidad Metropolitana de Ciencias de la Educación,}}\\
 \small{\textit{Av. José Pedro Alessandri 774, Ñuñoa, Santiago, Chile.}}\\
 \small{e-mail: jorge.pinochet@umce.cl}\\}

\date{} 
\maketitle

\begin{center}\rule{0.9\textwidth}{0.1mm} \end{center}
\begin{abstract}
\noindent El objetivo de este artículo es presentar una introducción no técnica a la física de los agujeros negros. Se discuten las principales propiedades de los cuatro tipos de agujero negro permitidos por el teorema de ausencia de pelo, y se examinan geométricamente algunas propiedades del espacio-tiempo alrededor de un agujero negro utilizando diagramas de inmersión.\\ 

\noindent \textbf{Descriptores}: Física de Agujeros Negros, Agujeros Negros Clásicos, Relatividad General Clásica.   

\end{abstract}

\selectlanguage{USenglish}

\begin{abstract}
\noindent The objective of this work is to present a non-technical introduction to black hole physics. The main properties of the four types of black hole allowed by the no-hair theorem are discussed, and some properties of spacetime around a black hole are geometrically examined using embedding diagrams. \\ 



\begin{center}\rule{0.9\textwidth}{0.1mm} \end{center}
\end{abstract}

\selectlanguage{spanish}

\maketitle

\section{Introducción}

Los agujeros negros, predichos por la teoría de la relatividad general hace más de un siglo, han fascinado durante mucho tiempo a especialistas y legos con sus sorprendentes propiedades, lo que sin duda ha contribuido a su enorme popularidad. La primera imagen de un agujero negro difundida en 2019 por la colaboración EHT, y el premio nobel de física 2020 otorgado a Roger Penrose, Reinhard Genzel y Andrea Ghez por sus importantes descubrimientos sobre los agujeros negros \cite{CUB}, han incrementado aún más la popularidad de estos objetos, despertando el interés de personas que nunca se habían acercado a este tema. Parece oportuno aprovechar este favorable escenario para invitar a un público lo más amplio posible a descubrir los secretos de los agujeros negros. \\

El objetivo de este trabajo es explorar algunos aspectos básicos de la física de los \textit{agujeros negros clásicos}, que son aquellos que se describen únicamente en el marco de la relatividad general. El artículo está dirigido a aquellos lectores que dominan los fundamentos de la física y el álgebra, y que están familiarizados con la física relativista.\\  

El tema que abordaremos es muy amplio, lo que nos obliga a ser selectivos. Por tanto, este trabajo no pretende ofrecer un análisis pormenorizado de la física de los agujeros negros, sino que solo busca convertirse en una breve introducción que sirva de guía y complemento para lecturas más profundas. Un tema importante que omitiremos son los agujeros negros cuánticos, que son aquellos cuya descripción requiere conciliar la relatividad general y la teoría cuántica. Para aquellos lectores que deseen ahondar en este tema, en dos artículos recientes el autor ha analizado las propiedades cuánticas de los agujeros negros \cite{PIN1, PIN2}.\\

El artículo está organizado del siguiente modo. En la primera parte se presenta una breve introducción al concepto de agujero negro clásico. Luego se analizan las propiedades de los distintos tipos de agujeros negros predichos por la relatividad general. Después se examinan geométricamente las propiedades del espacio-tiempo alrededor de un agujero negro. El artículo finaliza con unos breves comentarios. 

\section{Agujeros negros clásicos}

Los agujeros negros son un tipo de solución matemática exacta a las ecuaciones de la relatividad general, que es la teoría de la gravedad propuesta por Einstein en 1916 para perfeccionar y ampliar la ley de gravedad de Newton. Sin embargo, se ha encontrado que dicha solución matemática describe objetos que existen en el universo real, como el agujero negro supermasivo "fotografiado" por la colaboración EHT.\\ 

Podemos definir un agujero negro como una región del espacio-tiempo que contiene una concentración tan elevada de materia, que nada puede escapar de su gravedad, ni siquiera la luz \cite{LUM}. Un agujero negro es un objeto muy simple si se le compara, por ejemplo, con una enana blanca o una estrella de neutrones, con las cuales comparte el hecho de ser \textit{objetos compactos}, como se conoce a los cuerpos celestes caracterizados por una elevada concentración de materia.\\ 

Esta simplicidad se explica en base al \textit{teorema de ausencia de pelo}, un resultado de la relatividad general que establece que existen solo tres parámetros clásicos observables externamente que definen un agujero negro: masa $M$, carga eléctrica $Q$, y momentum angular $J$. Esto significa que clásicamente existen solo cuatro tipos de agujero negro, que llevan los nombres de los físicos que encontraron las correspondientes soluciones a las ecuaciones de la relatividad general \cite{LUM, FRO}: (1) el agujero negro de Schwarzschild, que solo depende de $M$; (2) el de Reissner-Nordstrom, que depende de $M$ y $Q$; (3) el de Kerr, que depende de $M$ y $J$; y el más general, (4) el de Kerr-Newman, que depende de $M$, $J$ y $Q$. En las siguientes secciones analizaremos por separado estos objetos. 

\section{Agujero negro de Schwarzschild}

La primera solución exacta a las ecuaciones de la relatividad general para lo que hoy en día denominamos agujero negro, fue encontrada por el astrónomo alemán Karl Schwarzschild en 1916 \cite{BART}. Para comprender intuitivamente el significado de esta solución matemática, imaginemos un objeto esférico de masa $M$ y radio $r$. En la medida que $r$ se reduce, la gravedad del objeto se incrementa más y más, hasta que llega un momento en que su densidad es tan grande, que ni siquiera la luz puede escapar de su gravedad. El valor crítico que debe tomar $r$ para que esto ocurra se denomina \textit{radio de Schwarzschild} y se calcula mediante la ecuación\footnote{Existe un argumento newtoniano intuitivo para obtener la ecuación (1). Si consideramos un objeto masivo de radio $R$ y masa $M$, la velocidad de escape desde su superficie viene dada por $V_{e} = (2GM/R)^{1/2}$. Si tomamos $V_{e} = c$, y resolvemos para $R$ se obtiene: $R = 2GM/c^{2}$. El significado físico de esta expresión es claro: ninguna forma de materia o energía contenida dentro de la superficie esférica limitada por $R$ puede escapar, ya que para ello necesitaría una rapidez mayor que $c$.} \cite{FRO}:
\begin{equation}\label{Ec1}
R_{S} = \frac{2GM}{c^{2}} \simeq 1,48 \times 10^{-27} m \left( \frac{M}{kg} \right), 
\end{equation}
donde $G = 6.67 \times 10^{-11} N\cdot m^{2} \cdot k^{-2}$ es la constante de gravitación y $c = 3 \times 10^{8} m \cdot s^{-1}$ es la rapidez de la luz en el vacío. Pero según la relatividad general, el objeto no puede mantenerse estático en su radio de Schwarzschild, pues la gravedad que ejerce sobre sí mismo es tan poderosa, que sobreviene un colapso que ninguna fuerza conocida puede detener. El colapso gravitacional termina cuando toda la masa queda reducida a un punto matemático de densidad infinita conocido como \textit{singularidad}, localizado en el centro de la esfera de radio $R_{S}$; por definición, en ese momento se ha formado un agujero negro. La esfera de radio $R_{S}$ se denomina \textit{horizonte} y define el límite que separa al agujero negro del resto del universo. Un observador externo solo puede ver lo que sucede fuera del horizonte. La Fig. 1 muestra la estructura interna de un agujero negro de Schwarzschild, también conocido como \textit{agujero negro estático}, donde se aprecia que tiene simetría esférica.\\

\begin{figure}[h]
\centering
    \includegraphics[width=0.3\textwidth]{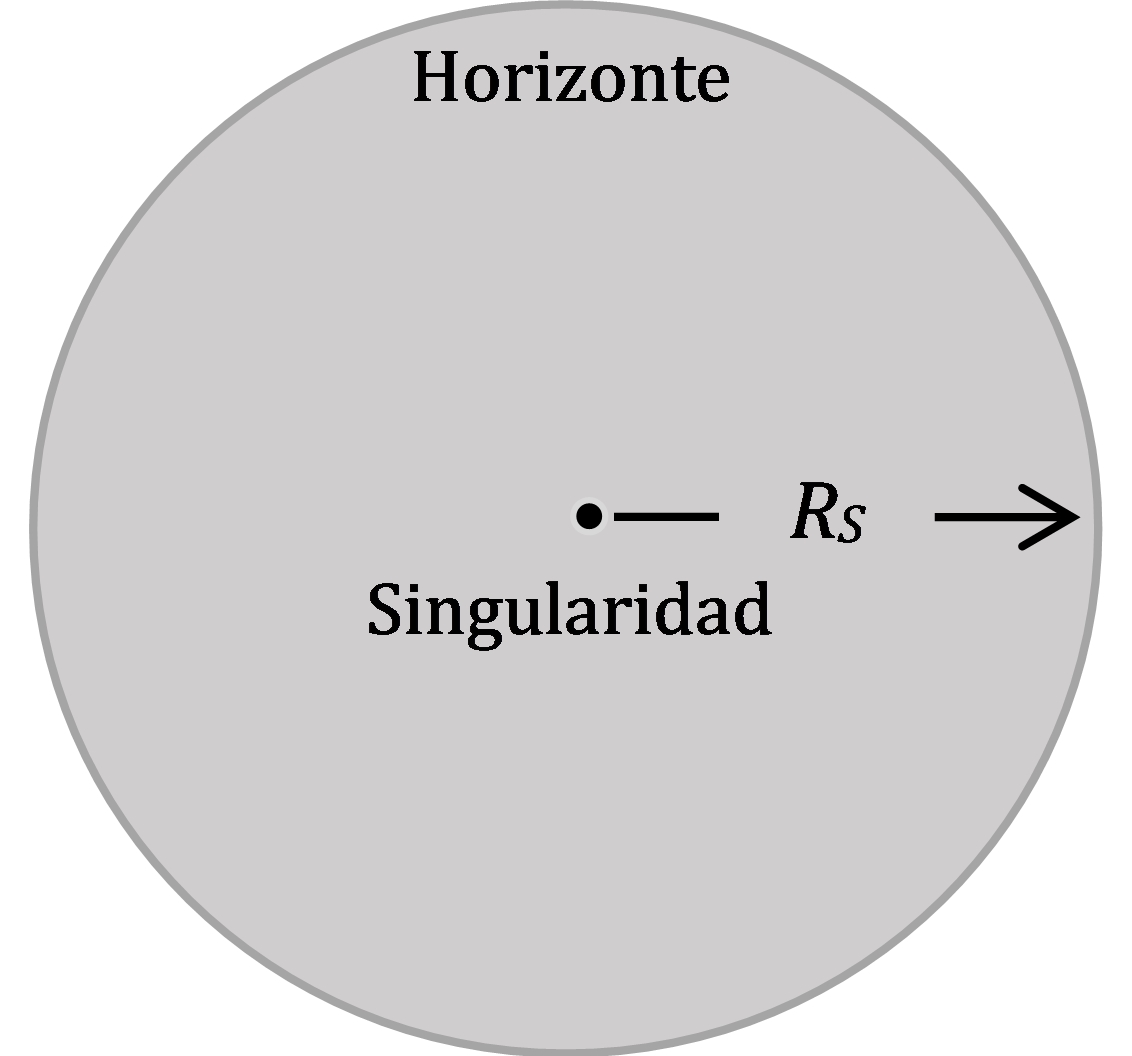}
\caption{\label{Fig1} Estructura interna de un agujero negro de Schwarzschild.}
\end{figure}

Un objeto que esté fuera del horizonte puede escapar de la gravedad del agujero negro si tiene suficiente energía, pero si se encuentra dentro del horizonte, quedará causalmente desconectado del exterior, e inevitablemente será absorbido por la singularidad. Aunque el horizonte no tiene existencia material, puede imaginarse como una membrana unidireccional que solo permite el flujo de materia y energía del exterior al interior, pero nunca en sentido contrario.\\ 

Para tener una idea de las colosales concentraciones de masa, o de su equivalente en energía, implicadas en la formación de un agujero negro, notemos que si en la ecuación (1) tomamos la masa terrestre, $M = 5,97 \times 10^{24} kg$ se obtiene $R_{S} \simeq 9 mm$ que es el tamaño de una canica.\\ 

Es importante hacer notar que si bien, de acuerdo con la Fig. 1, $R_{S}$ representa la distancia radial desde la singularidad hasta el horizonte, cerca de un agujero negro y en su interior las distancias radiales son solo parámetros y no tienen el significado físico que les atribuimos en nuestra experiencia directa. Por tanto, la Fig. 1 solo tiene un valor pedagógico y no debe considerarse una representación exacta de la realidad. Esto también es válido para las figuras que usaremos en las siguientes secciones para representar la estructura interna de los otros tipos de agujero negro. Estas ideas quedarán más claras cuando lleguemos a la Sección VII.

\section{Agujero negro de Reissner-Nordstrom}

Poco tiempo después de que Schwarzschild encontrara la solución que lleva su nombre, los físicos Hans Reissner y Gunnar Nordstrom descubrieron, en forma independiente, una solución a la ecuación de Einstein que representa el espacio tiempo de un agujero negro cargado eléctricamente \cite{LUM, FRO}. La estructura de un agujero de Reissner-Nordstrom es más compleja que la de un agujero de Schwarzschild, pues tiene dos horizontes esféricos concéntricos, uno interno y otro externo. Al igual que en la solución de Schwarzschild, un observador en el exterior solo puede ver lo que sucede fuera del horizonte externo. Para un agujero de Reissner-Nordstrom con carga $Q$, el radio del horizonte externo es \cite{FRO}: 
\begin{equation}\label{Ec2}
R_{+} = \frac{GM}{c^{2}} + \sqrt{\frac{G^{2}M^{2}}{c^{4}} - \frac{GQ^{2}}{4\pi \varepsilon_{0}c^{4}}},
\end{equation}
donde $\varepsilon_{0} = 8,85 \times 10^{-12} C^{2}\cdot N \cdot m^{-2}$ es la permitividad del vacío. El radio del horizonte interno es: 
\begin{equation}\label{Ec3}
R_{-} = \frac{GM}{c^{2}} - \sqrt{\frac{G^{2}M^{2}}{c^{4}} - \frac{GQ^{2}}{4\pi \varepsilon_{0}c^{4}}}.
\end{equation}
Se aprecia que $R_{+} \geq R_{-}$. Estas dos ecuaciones pueden sintetizarse como: 
\begin{equation}\label{Ec4}
R_{\pm} = \frac{R_{S}}{2} \pm \sqrt{\frac{R_{S}^{2}}{4} - \frac{GQ^{2}}{4\pi \varepsilon_{0}c^{4}}},
\end{equation}
donde hemos introducido la ecuación (1). Para $Q = 0$, el agujero de Reissner-Nordstrom se convierte en uno de Schwarzschild, ya que $R_{-} = 0$ y $R_{+} = R_{S}$. Al igual que en la solución de Schwarzschild, la singularidad central es un punto matemático. La Fig. 2 representa la estructura interna de un agujero negro de Reissner-Nordstrom. Se aprecia que este objeto tiene simetría esférica.\\  

\begin{figure}[h]
\centering
    \includegraphics[width=0.4\textwidth]{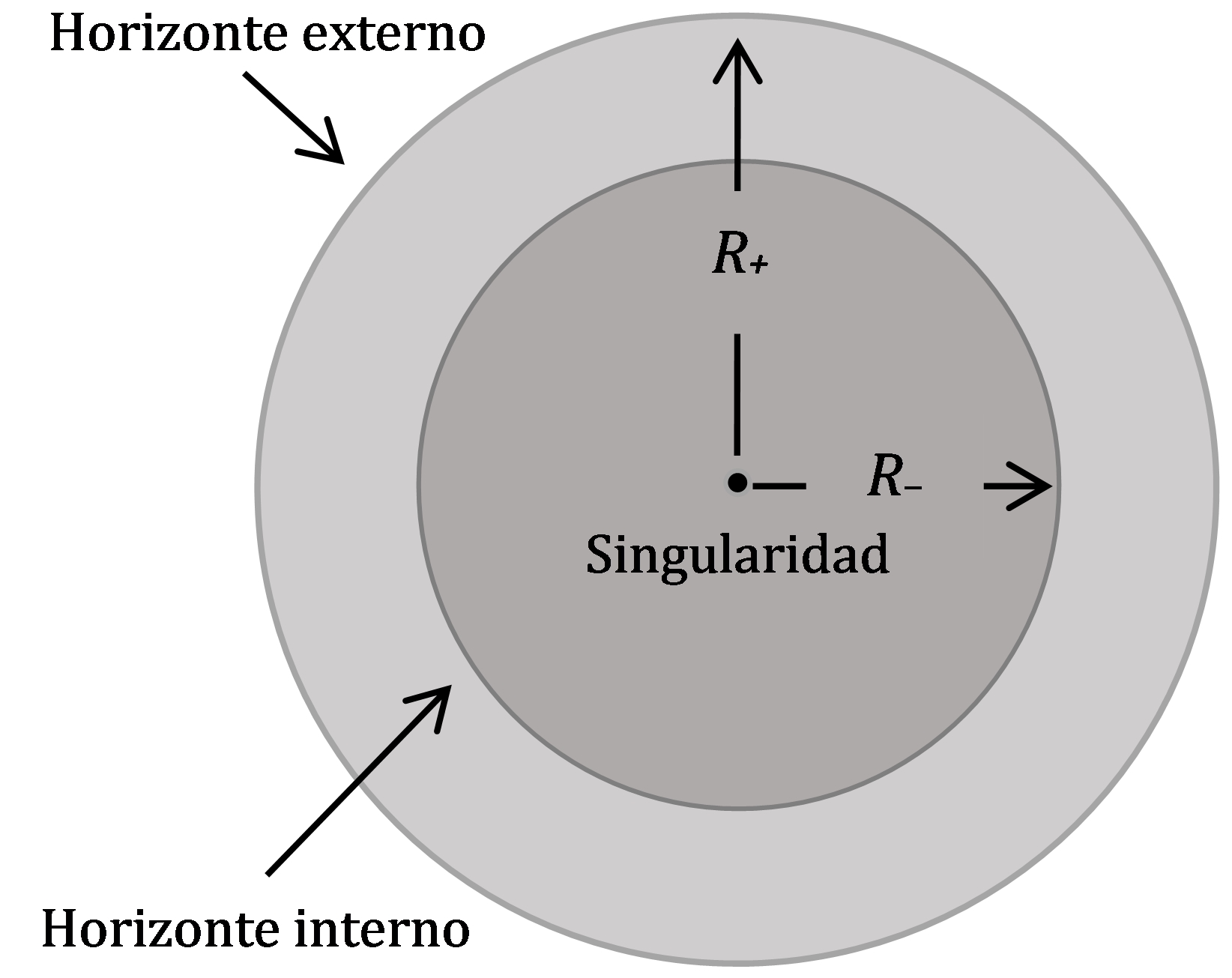}
\caption{\label{Fig2} Estructura interna de un agujero negro de Reissner-Nordstrom.}
\end{figure}

Intuitivamente es evidente que $Q$ no puede aumentar indefinidamente, pues la carga provoca un efecto repulsivo que se opone a la gravedad. Si la repulsión eléctrica superara a la gravedad, el agujero negro desaparecería, y en su lugar se formaría una singularidad desnuda, carente de horizonte. Por razones teóricas, se presume que no pueden existir singularidades desnudas. Esta conjetura se denomina \textit{censura cósmica}, y fue formulada por Penrose en 1969.\\ 

A partir de la ecuación (4) podemos calcular la carga máxima que puede tener un agujero de Reissner-Nordstrom. Como la cantidad subradical no puede hacerse negativa, la carga máxima se calcula haciendo que dicha cantidad se anule: 
\begin{equation}\label{Ec5}
Q_{max} = 2\sqrt{\pi \varepsilon_{0}G} M \simeq 8,6 \times 10^{-11} C \left( \frac{M}{kg} \right).  
\end{equation}
Cuando se alcanza esta carga, se habla de un agujero negro de Reissner-Nordstrom \textit{extremal}. Se presume que los agujeros negros astrofísicos deben tener una carga neta nula o muy pequeña, ya que un agujero con $Q \neq 0$ rápidamente atraerá cargas opuestas que lo llevarán a la neutralidad. En este sentido, los agujeros negros de Reissner-Nordstrom revisten un interés más matemático que astrofísico. 

\section{Agujero negro de Kerr}

Casi medio siglo después de los descubrimientos de Schwarzschild, Reissner y Nordstrom, el físico neozelandés Roy Kerr encontró una solución exacta para el espacio-tiempo de un agujero negro con momentum angular (rotación) \cite{LUM, FRO}. El agujero negro de Kerr es el más realista, ya que todos los objetos del universo tienen algún grado de rotación. No obstante, cuando la rapidez angular es pequeña, el agujero de Schwarzschild puede ser una buena aproximación para los agujeros negros astrofísicos. \\

Al igual que el caso analizado en la sección anterior, el agujero negro de Kerr tiene dos horizontes esféricos concéntricos, uno interno y otro externo (ver Figs. 3 y 4). Nuevamente, un observador en el exterior solo puede ver lo que sucede fuera del horizonte externo. Para un agujero de Kerr con momentum angular $J$, el radio del horizonte externo se escribe como \cite{FRO}: 
\begin{equation}\label{Ec6}
R_{+} = \frac{GM}{c^{2}} + \sqrt{\frac{G^{2}M^{2}}{c^{4}} - \frac{J^{2}}{c^{2}M^{2}}}
\end{equation}
El radio del horizonte interno es: 
\begin{equation}\label{Ec7}
R_{-} = \frac{GM}{c^{2}} - \sqrt{\frac{G^{2}M^{2}}{c^{4}} - \frac{J^{2}}{c^{2}M^{2}}}
\end{equation}
Vemos que $R_{+} \geq R_{-}$. Estas dos ecuaciones pueden sintetizarse como: 
\begin{equation}\label{Ec8}
R_{\pm} = \frac{R_{S}}{2} \pm \sqrt{\frac{R_{S}^{2}}{4} - \frac{J^{2}}{c^{2}M^{2}}}.
\end{equation}
Se observa que para $J = 0$, el agujero de Kerr se convierte en estático, ya que $R_{-} = 0$ y $R_{+} = R_{S}$. La estructura de un agujero negro de Kerr es más compleja que las analizadas antes. En primer lugar, la singularidad central no es un punto sino un anillo localizado sobre el plano ecuatorial. En segundo lugar, además de sus dos horizontes concéntricos, tiene una región que se encuentra fuera del horizonte exterior llamada \textit{ergósfera}, donde ningún objeto puede mantenerse estático (Fig. 3). De acuerdo con el \textit{efecto Lense–Thirring}, el espacio-tiempo en las proximidades del agujero de Kerr es arrastrado por la rotación de éste, arrastrando consigo a todos los objetos situados en la ergósfera, impidiendo que permanezcan estáticos\footnote{Un agujero de Kerr se encuentra en rotación perfectamente rígida; todos los puntos del horizonte tienen la misma rapidez angular.}. Por esta razón, al límite exterior de la ergósfera se le conoce como \textit{límite estático}. Como la ergósfera está fuera del horizonte, un objeto que se desplace suficientemente rápido hacia el exterior puede escapar del agujero negro, siempre que se encuentre fuera del horizonte externo.\\ 

El radio del límite estático se calcula como \cite{FRO}: 
\begin{equation}\label{Ec9}
r = \frac{R_{S}}{2} + \sqrt{\frac{R_{S}^{2}}{4} - \frac{J^{2}}{M^{2}c^{2}}cos^{2}\theta},
\end{equation}
donde $\theta$ es el ángulo polar (Fig. 5). Esta ecuación muestra que en los polos (donde $\theta = 0^{o}$ y $\theta = 180^{o}$) $r = R_{+}$, y en el ecuador (donde $\theta = 90^{o}$) $r = R_{S}$. 

\begin{figure}[h]
\centering
    \includegraphics[width=0.4\textwidth]{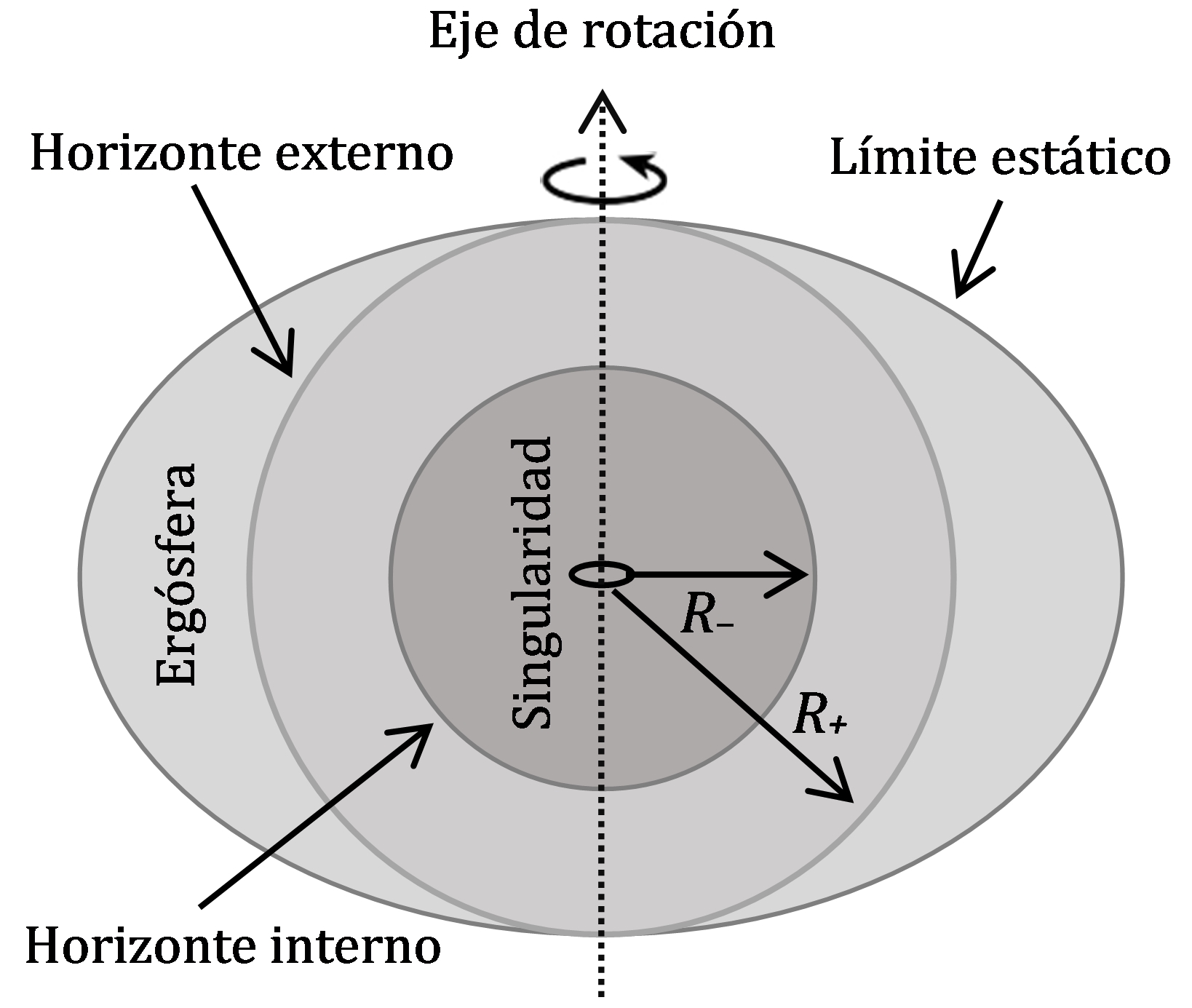}
\caption{\label{Fig3} Agujero negro de Kerr visto paralelamente a su eje de giro. El ensanchamiento de la ergósfera en el ecuador se debe a la rotación.}
\end{figure}

\begin{figure}[h]
\centering
    \includegraphics[width=0.5\textwidth]{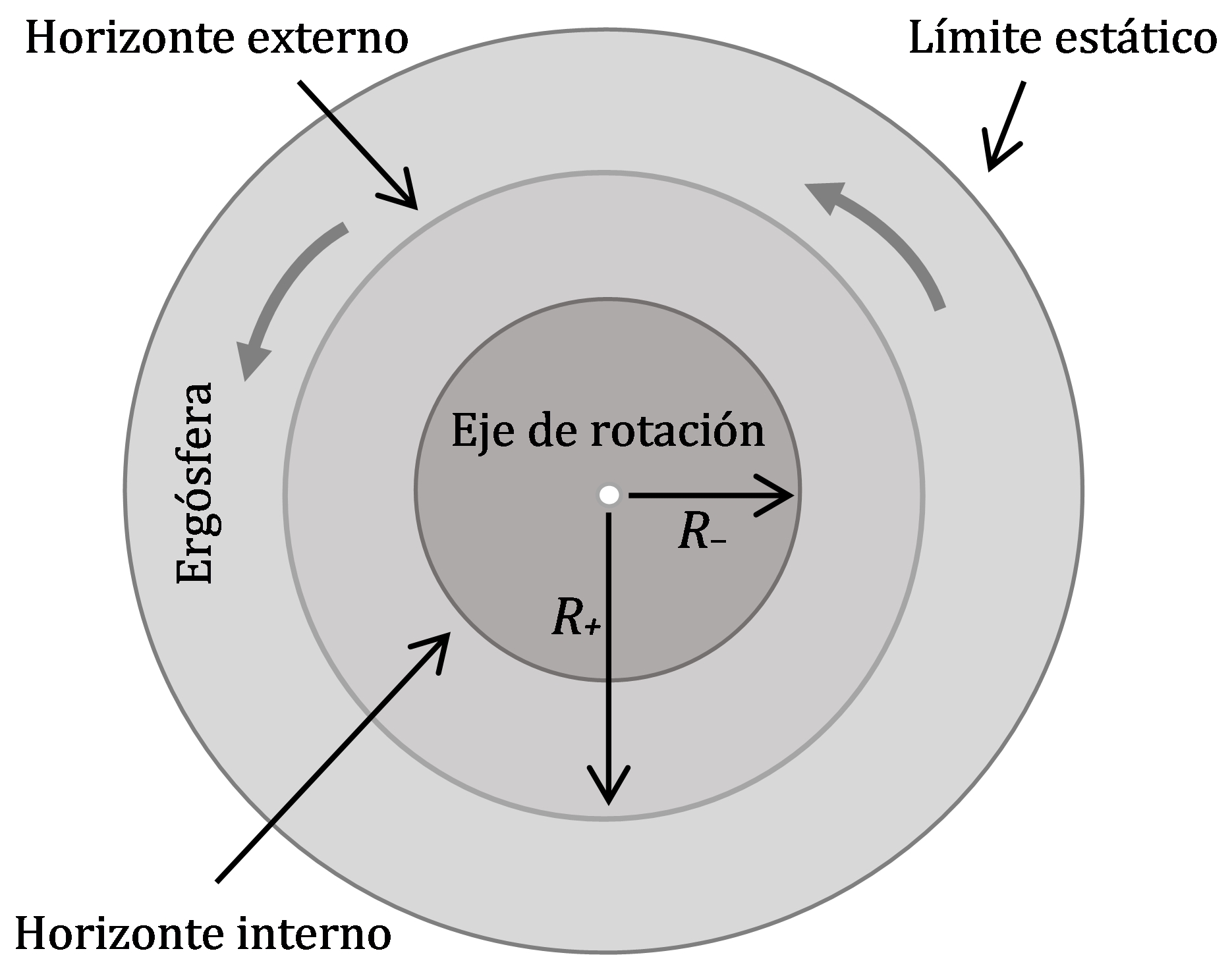}
\caption{\label{Fig4} Agujero negro de Kerr visto desde uno de sus polos.}
\end{figure}

Como la rotación es una forma de energía cinética, los agujeros de Kerr pueden almacenar una gran cantidad de energía en la ergósfera, y como esta región se encuentra fuera del horizonte externo, puede ser extraída. Se han propuesto diversos mecanismos para efectuar dicha extracción. Uno de los más populares es el \textit{proceso de Penrose}, propuesto por Penrose en 1967.\\ 

Cálculos detallados revelan que la energía máxima que puede extraerse de un agujero de Kerr es un 29\% de su masa inicial. Para tener una base de comparación, recordemos que la eficiencia de una reacción de fusión nuclear es $\sim 1\%$. Una vez que se ha extraído toda la energía rotacional, el agujero de Kerr deja de girar y se convierte en estático, como se demuestra al tomar $J = 0$ en la ecuación (8). Cuando ello ocurre, clásicamente no es posible extraer más energía del agujero de Kerr.\\ 

\begin{figure}[h]
\centering
    \includegraphics[width=0.5\textwidth]{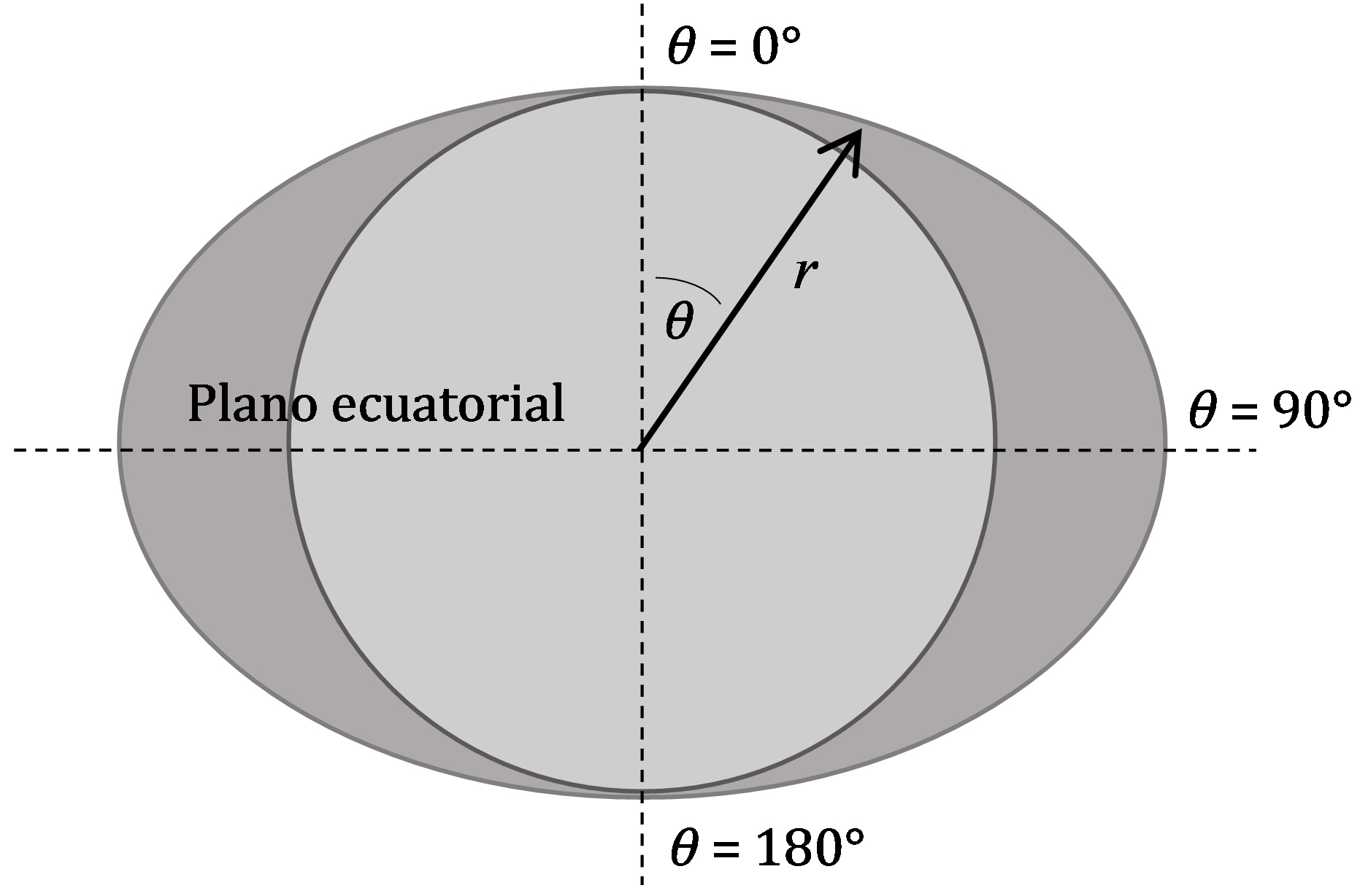}
\caption{\label{Fig5} Radio del límite estático en función del ángulo polar $0^{o} \leq \theta \leq 180^{o}$.}
\end{figure}

Es evidente que $J$ no puede aumentar indefinidamente, pues la rotación genera una fuerza centrífuga que se opone a la gravedad. Si la fuerza centrífuga superara a la gravedad, el agujero negro desaparecería y se formaría una singularidad desnuda. Para calcular el valor máximo de $J$ anulamos la cantidad subradical en la ecuación (8):
\begin{equation}\label{Ec10}
J_{max} = \dfrac{1}{2}McR_{S} = \frac{GM^{2}}{c} \simeq 2,2 \times 10^{-19} kg \cdot m^{2} \cdot s^{-1} \left( \frac{M}{kg} \right)^{2}.  
\end{equation} 
Cuando se alcanza este momentum angular, se habla de un agujero de Kerr extremal. En este estado, la rapidez de rotación del horizonte es cercana a $c$. Los agujeros de Kerr permiten modelar los fenómenos más energéticos del universo, como los \textit{núcleos activos de galaxias}, cuyas fantásticas emisiones de energía se explican mediante un modelo donde un agujero negro supermasivo qua está próximo al estado extremal, actúa como un motor gravitacional cuyo combustible es un disco de acreción.

\section{Agujero negro de Kerr-Newman}
Dos años después de que Kerr encontrara la solución que lleva su nombre, el físico Ezra T. Newman y sus colaboradores descubrieron una solución exacta a la ecuación de Einstein que representa el espacio tiempo de un agujero negro cargado y rotante \cite{LUM, FRO}. Debido a la presencia de carga, esta solución tiene escaso interés astrofísico.\\  

En analogía con las soluciones de Reissner-Nordstrom y de Kerr, el agujero negro de Kerr-Newman posee dos horizontes concéntricos cuyos radios son \cite{FRO}: 
\begin{equation}\label{Ec11}
R_{\pm} = \frac{GM}{c^{2}} \pm \sqrt{\frac{G^{2}M^{2}}{c^{4}} - \frac{J^{2}}{c^{2}M^{2}} - \frac{GQ^{2}}{4\pi \varepsilon_{0}c^{4}}}.
\end{equation}
La condición para que $Q$ y $J$ combinados tengan el valor máximo permitido se obtiene anulando la cantidad subradical: 
\begin{equation}\label{Ec12}
\frac{G^{2}M^{2}}{c^{4}} = \frac{J^{2}}{c^{2}M^{2}} + \frac{GQ^{2}}{4\pi \varepsilon_{0}c^{4}}.
\end{equation}
Fuera de esto, el agujero de Kerr-Newman posee cualitativamente la misma estructura que el de Kerr. 

\section{La geometría del espacio-tiempo en torno a un agujero negro}
Según la relatividad general, la gravedad no es una fuerza como supuso Newton, sino una manifestación de la curvatura del espacio-tiempo. Dicho espacio-tiempo tiene tres dimensiones espaciales y una temporal, de modo que es cuatridimensional (4D). Los objetos que se mueven libremente en este espacio-tiempo, se limitan a seguir las trayectorias dictadas por la geometría espaciotemporal. En este sentido, la masa le dice al espacio-tiempo como curvarse, y el espacio-tiempo le dice a la masa como moverse \cite{WHEE}.\\ 

En las cercanías de un agujero negro, el espacio-tiempo se encuentra muy curvado. Un recurso utilizado frecuentemente para representar esta curvatura es el \textit{diagrama de inmersión} \cite{STE}. Para estudiar el espacio-tiempo alrededor de un agujero negro mediante este tipo de diagrama, podemos centrar nuestra atención en la geometría de Schwarzschild, que es la más simple. Cualitativamente, el espacio-tiempo en torno a los otros tipos de agujero negro es similar a la de Schwarzschild.\\ 

Para comenzar, debemos tener presente que la geometría de Schwarzschild es estática y estacionaria (no depende del tiempo), lo cual significa que no se pierde ninguna información si consideramos el tiempo constante; además, como dicha geometría es esféricamente simétrica, tampoco se pierde información si solo consideramos planos paralelos al plano ecuatorial. Lo anterior equivale formalmente a eliminar dos dimensiones, una temporal y otra espacial. El beneficio de esto es que la geometría de Schwarzschild queda reducida a una superficie curva 2D que podemos visualizar, ya que queda inmersa (de aquí proviene el nombre "diagrama de inmersión") en un espacio 3D. Esta superficie 2D se denomina \textit{paraboloide de Flamm}, y se genera por la rotación continua de una parábola en torno a un eje perpendicular al plano ecuatorial (eje $z$). La ecuación de esta parábola es \cite{STE}: 
\begin{equation}\label{Ec13}
z = \pm 2\sqrt{R_{S}(r-R_{S})} = \pm 2\sqrt{R_{S}(\sqrt{x^{2}+y^{2}}-R_{S})}.
\end{equation}
La Fig. 6 muestra la parábola en color negro, donde se muestra la misma parábola, en rojo, rotada $180^{o}$ respecto del eje $z$. La Fig. 7 muestra el paraboloide de Flamm obtenido al efectuar una rotación de la parábola en torno al eje $z$, donde solo se muestra la región $z^{+}$. El plano ecuatorial corresponde al plano $x-y$. Vemos que la ecuación (13) también admite una solución negativa, correspondiente a la región $z^{-}$. Esta región se relaciona con los agujeros de gusano, un tema fascinante y controvertido que no analizaremos en este trabajo. El lector interesado puede encontrar una excelente discusión no técnica en \cite{LUM}. Un análisis más técnico se encuentra en \cite{THORNE1}. \\

\begin{figure}[h]
\centering
    \includegraphics[width=0.6\textwidth]{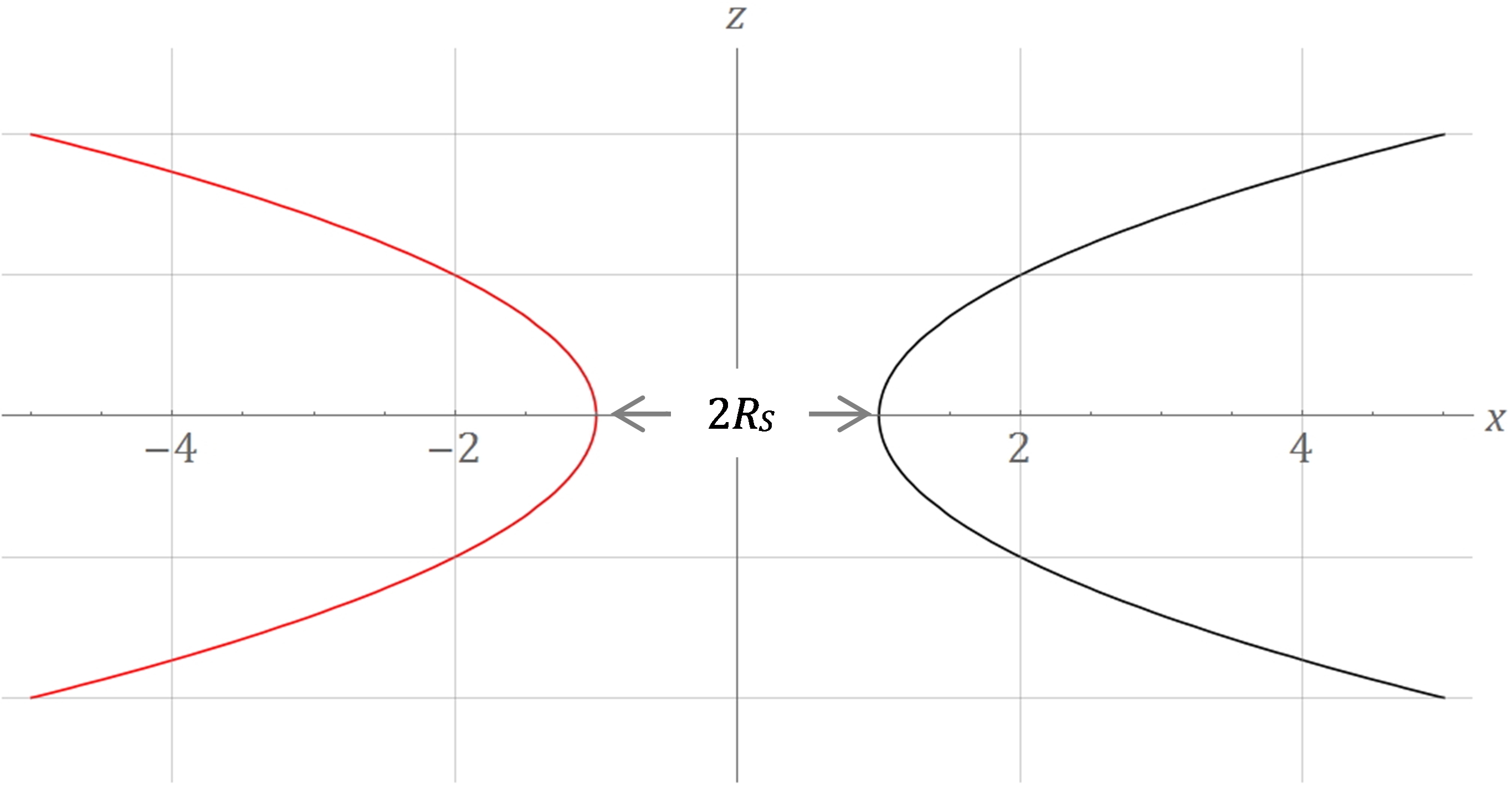}
\caption{\label{Fig6} La rotación de la parábola negra en torno al eje z genera el espacio-tiempo de Schwarzschild 2D.}
\end{figure}

Para obtener las gráficas hemos tomado $R_{S} = 1$, de modo que los ejes $x$ e $y$ están expresados en unidades del radio de Schwarzschild. Se observa que la parábola corta el eje $x$ en $R_{S}$. Notemos que el diagrama no muestra el interior del agujero negro ni la singularidad. La razón es que para $r < R_{S}$, el lado derecho de la ecuación (13) se vuelve complejo.\\

La forma del paraboloide recuerda la analogía de la tela elástica (rubber sheet analogy), utilizada frecuentemente en discusiones elementales para representar la curvatura del espacio-tiempo. Sin embargo, es importante tener presente que la similitud entre ambas representaciones es superficial. La principal diferencia es que solo el paraboloide de Flamm es matemáticamente riguroso. Además, recordemos que la construcción del paraboloide supone que el tiempo es constante, lo que significa que la Fig. 7 es una suerte de fotografía del espacio-tiempo; en consecuencia, a diferencia de lo que ocurre con la analogía de la tela elástica, no pueden existir objetos moviéndose sobre la superficie del paraboloide. \\

Las Figs. 6 y 7 muestran que las distancias radiales sobre el paraboloide son mayores que las correspondientes distancias sobre el plano $x-y$. Es decir, las distancias radiales en el espacio-tiempo curvado por un agujero negro son mayores que las correspondientes distancias en un espacio-tiempo plano. Por ejemplo, si en la Fig. 6 consideramos los puntos $x = 2$ y $x = 4$, vemos que la distancia entre estos puntos medida sobre el eje $x$ es menor que la correspondiente distancia medida sobre la parábola (curva negra). Todo esto revela que en las proximidades de un agujero negro no podemos atribuir a las distancias radiales un significado físico directo; dichas distancias se miden sobre un espacio fuertemente curvado (representado por la superficie del baraboloide cerca de $r = R_{S}$) cuya geometría es no euclidiana, mientras que las distancias de nuestra experiencia cotidiana pueden medirse con un alto grado de aproximación mediante la geometría plana de Euclides, pues la gravedad que experimentamos en nuestro entorno es muy débil, y los efectos de la relatividad general y de la curvatura espaciotemporal pueden despreciarse.\\

\begin{figure}[h]
\centering
    \includegraphics[width=0.6\textwidth]{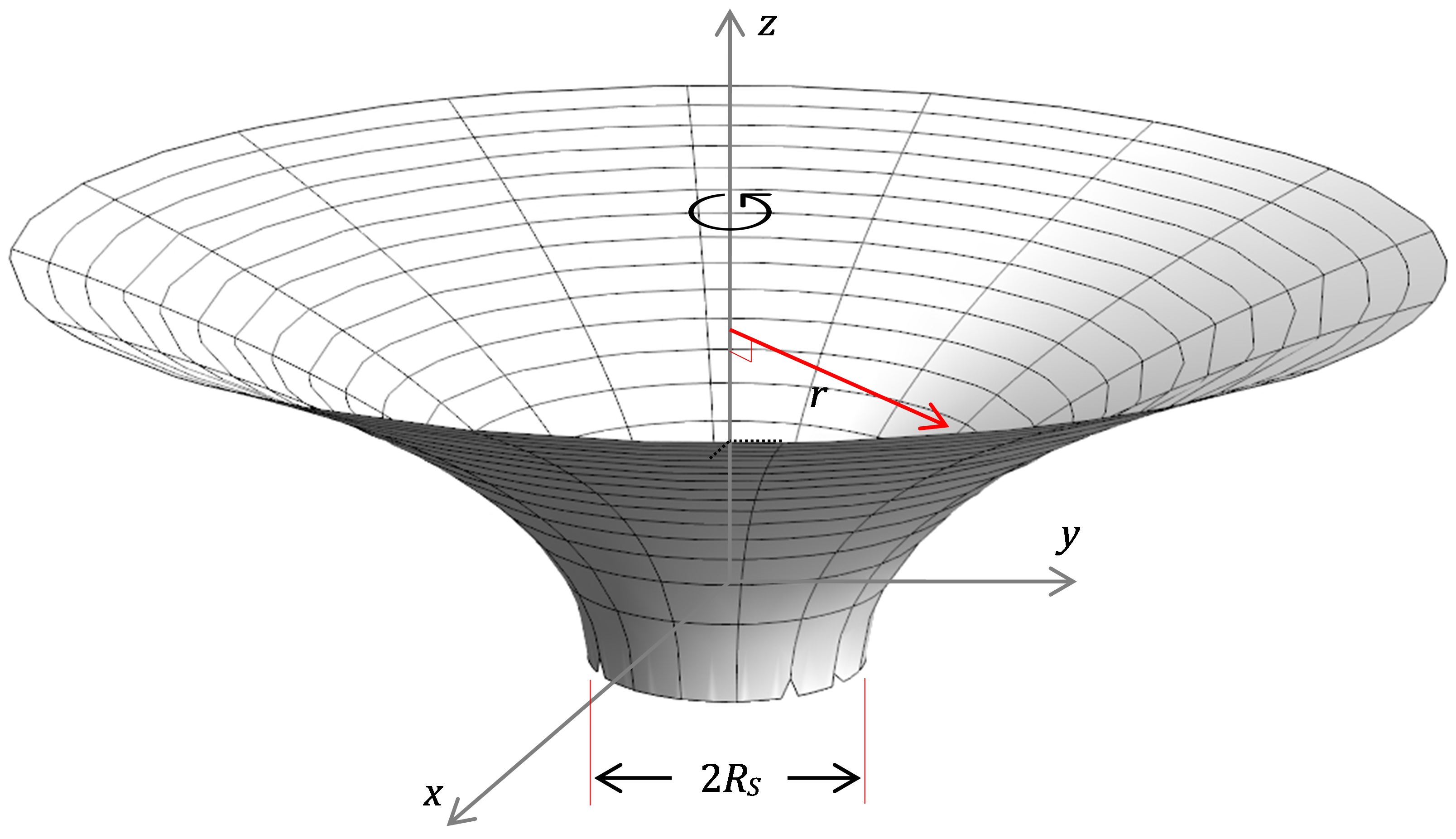}
\caption{\label{Fig7} El paraboloide de Flamm representa el espacio-tiempo de Schwarzschild 2D.}
\end{figure}

Otro aspecto importante de la geometría de Schwarzschild, que se aprecia en la Fig. 6, y que está directamente relacionado con lo discutido en el párrafo anterior, es que en la medida que nos alejamos del eje $z$, donde se localiza el agujero negro, la superficie de la parábola es cada vez más plana\bibliographystyle{Aquí estamos empleando una definición intuitiva de curvatura, entendida como una medida de la desviación que presenta un objeto geométrico respecto de la planitud o linealidad.} (menos curva), lo que implica que la superficie del paraboloide también es cada vez más plana. Técnicamente se dice que el espacio-tiempo es \textit{asintóticamente plano}. Podemos verificar esta propiedad derivando la ecuación (13):
\begin{equation}\label{Ec14}
\dfrac{dz}{dr} = \frac{R_{S}}{\sqrt{R_{S}(r-R_{S})}} = \frac{1}{\sqrt{r/R_{S} - 1}}.
\end{equation}
Esta expresión nos permite calcular la pendiente sobre la dirección radial en cualquier punto de la superficie del paraboloide. Vemos que cuando $r \rightarrow \infty$, $dz/dr \rightarrow 0$. Es decir, para $r \gg R_{S}$ la pendiente es nula y la superficie del paraboloide es plana. Esto significa que lejos del agujero negro los efectos de la relatividad general son cada vez más pequeños, y la ley de gravitación de Newton se convierte en una buena aproximación para describir la gravedad.\\

Finalmente es importante considerar que solo la superficie 2D del paraboloide de Flamm tiene significado como parte de la geometría de Schwarzschild. Los puntos fuera de la superficie carecen de significado físico, ya que en un diagrama de inmersión el universo queda reducido a la superficie 2D del paraboloide, lo que significa que el espacio 3D exterior no forma parte del universo. 

\section{Conclusiones}

Existen muchos temas sobre la física de los agujeros negros que por motivos de espacio no hemos abordado en este trabajo, como la espaguetización, los agujeros de gusano, o la termodinámica de los agujeros negros, entre otros. Quienes deseen profundizar en estos y otros temas afines, pueden recurrir a la extensa literatura, tanto divulgativa \cite{LUM, BART, HAC, THORNE2, SUS} como técnica \cite{FRO, RUF, BA, CHOW}. El lector también puede revisar otros artículos del autor sobre este tema \cite{PIN1, PIN2, PIN3, PIN4}. En cualquier caso, tengo la esperanza de que los tópicos escogidos sean una estimulante introducción a la física de los agujeros negros, y motiven al lector a continuar profundizando en este fascinante tema.

\section{Agradecimientos}

Quisiera agradecer a Daniela Balieiro por sus útiles comentarios durante la redacción de este artículo.

\end{document}